\documentclass{aa}

\usepackage{graphicx}

\usepackage{amsmath,amssymb}
\usepackage{txfonts}
\usepackage{natbib}
\usepackage{subfigure}

\usepackage{array}

\usepackage[normalem]{ulem}  
\usepackage{xcolor}

\definecolor{mygreen}{rgb}{0.0, 0.76, 0.0}

\usepackage{hyperref}
\usepackage{nameref}

\title{Morphological Evolution of Higher Order Nonlinear Kinetic Alfvén Waves in Structured Galactic Environments}

\titlerunning{Higher-Order KA Solitons in Structured ISM}

\author{
    Manpreet Singh\inst{1,2}\and
    Siming Liu\inst{2}\and
    N. S. Saini\inst{3}
}
\authorrunning{M. Singh et al.}

\institute{
    School of Computing and Artificial Intelligence, Southwest Jiaotong University, Chengdu-610031, PR China.\\
    \email{singhmanpreet185@gmail.com}
    \and
    School of Physical Science and Technology, Southwest Jiaotong University, Chengdu-610031, PR China.
    \and
    Department of Physics, Guru Nanak Dev University, Amritsar-143005, India
}

\date{Received <date>; accepted <date>}

\abstract
{Kinetic Alfvén waves (KAWs) are fundamental to energy transport and small‑scale structure formation in the turbulent, magnetized interstellar medium (ISM). While first‑order Korteweg–de Vries (KdV) models describe weakly nonlinear KAW solitons, they fail in strongly inhomogeneous environments where higher‑order dispersive and nonlinear effects become significant.}
{We investigate the existence, morphology, and Galactic distribution of higher‑order “dressed” kinetic Alfvén (KA) solitons in a structured ISM that includes the warm ionized medium, H\,\textsc{II} regions, stellar‑wind bubbles (SWBs), and supernova remnants (SNRs).} 
{Using a multi‑component fluid model with superthermal electron distributions and spatially varying background profiles, we derive an inhomogeneous KdV‑type equation that incorporates cubic nonlinearity, nonlinear‑dispersive cross terms, and fifth‑order dispersion. The dressed soliton solution consists of a leading‑order $\operatorname{sech}^2$ core decorated by higher‑order corrections. We classify soliton morphologies and map them across the Galactic plane as a function of the electron suprathermality index $\kappa_e$.}
{The morphology maps reveal five distinct soliton classes ($\psi_{\rm I}$–$\psi_{\rm V}$) whose distribution evolves non‑monotonically with $\kappa_e$. Strongly suprathermal plasmas ($\kappa_e=1.6$) favour negative double‑hump solitons ($\psi_{\rm III}$); intermediate $\kappa_e$ produce radially layered sequences of positive double‑hump ($\psi_{\rm II}$), single‑hump ($\psi_{\rm I}$), dressed ($\psi_{\rm IV}$), and split‑core ($\psi_{\rm V}$) states; near‑Maxwellian plasmas ($\kappa_e=3.1$) revert to KdV‑like positive single‑hump solitons over most of the disk. Localised $\psi_{\rm V}$ structures appear as a red ring around the SWB shell and a compact red core inside the SNR, demonstrating that embedded ISM structures actively generate distinct higher‑order soliton morphologies absent in a homogeneous medium. Radial amplitude diagnostics show that inside the SNR the central pulsar wind nebula is an exclusion zone, with negligible soliton amplitude immediately outside it.}
{First‑order KdV theory is insufficient in large parts of the ISM; dressed solitons are the natural nonlinear states. The ISM morphology selects the soliton class by modulating the competition between leading‑order and higher‑order terms. Localised $\psi_{\rm V}$ features provide a direct link between macroscopic ISM structures and coherent kinetic‑scale fluctuations, offering candidates for extreme scattering events and pulsar scintillation. The non‑monotonic dependence on $\kappa_e$ opens a path to constrain the electron suprathermality from scintillation observations. The framework provides a physics‑based sub‑grid model for galaxy‑scale simulations.}


\begin{document}

\maketitle

\section{Introduction}

The interstellar medium (ISM) is a weakly collisional, magnetized plasma characterized by broadband turbulence and coherent electromagnetic fluctuations.~At kinetic scales, where the assumption of ideal magnetohydrodynamics breaks down, Alfvénic fluctuations develop finite parallel electric fields and compressibility, transitioning into Kinetic Alfvén Waves (KAWs) \citep{HASEGAWAParticle1976,CHASTONKinetic2003}. These modes are fundamental to energy transport in astrophysical plasmas, facilitating particle heating, energization, and the formation of localized nonlinear structures \citep{Stasiewicz2000,CHASTONKinetic2003,Howes2008JGR,Schekochihin2009,CHANDRANPERPENDICULAR2010}. Due to the delicate balance between dispersion and nonlinearity, KAWs can evolve into stable solitary wave structures, which are typically described using reductive perturbation theory and Korteweg--de Vries (KdV)–type evolution equations \citep{Bains2014}.

In a recent theoretical development, \citet{SINGHInterstellar2026} (hereafter Singh2026) introduced a spatially structured model of the ionized ISM to investigate Kinetic Alfvén (KA) soliton propagation. By incorporating realistic Galactic features, such as stratified density profiles at H\,\textsc{ii} regions, stellar wind bubbles (SWBs) and supernova remnants (SNRs), that work derived a first-order KdV equation governing soliton evolution. The study successfully demonstrated that the amplitude and width of KA solitons are intrinsic functions of the local plasma environment. However, the standard first-order KdV approximation employed in Singh2026 and similar studies is formally valid only in the limit of weak nonlinearity and weak dispersion.
In realistic astrophysical environments, these perturbative assumptions are frequently violated. Steep spatial gradients, mesoscale inhomogeneities, and large-amplitude fluctuations can drive the system into regimes where higher-order nonlinear and dispersive effects are no longer negligible. 
It has long been recognized in laboratory experiments \citep{watanabe1993} and theoretical treatments \citep{Kodama1978} that first-order theories fail to accurately describe larger amplitude structures or those propagating with phase speeds close to the linear wave speed of the underlying dispersive mode.
In such regimes, the inclusion of higher-order corrections is essential and leads to the formation of \emph{dressed} solitons, structures characterized by a localized core modified by asymmetric halos or oscillatory wave trains \citep{Sugimoto1977}.

Considerable effort has been devoted to characterizing these higher-order structures in various plasma environments. Extensive studies in electrostatic wave regimes have demonstrated that higher-order corrections significantly modify the soliton velocity, width, and existence domains, often resulting in non-standard profiles such as flat-top or negative-potential solitary waves \citep{Lab93,She07,Meh12,Sunidhi21,Raj22,Geetika22}. In the context of magnetized plasmas, \citet{MAHMOODHigherorder2021} and \citet{Kaur2023} have explored higher-order corrections to inertial and kinetic Alfvén modes respectively, revealing that these dressed structures can support quasi-periodic wave trains and, under specific conditions, transition into chaotic dynamics \citep{Mahmood20,Samanta13,Zhen14}.

The morphology of these nonlinear structures is strongly regulated by the energetic state of the particle distribution.~Observations consistently indicate that space and astrophysical plasmas often depart from Maxwellian equilibrium, exhibiting high-energy suprathermal tails well-modeled by Kappa distributions \citep{VASYLIUNASSurvey1968,Mak97}. Previous investigations have established that suprathermality significantly alters the nonlinearity coefficients of the governing evolution equations, thereby modifying the amplitude and width of solitons \citep{Bal08,Sai08}.

Despite this extensive literature, a critical gap remains: the behavior of dressed KA solitons has not yet been self-consistently modeled within a realistically structured astrophysical medium. Most existing studies rely on idealized, homogeneous plasma models that neglect the systematic spatial variation of plasma parameters (density, temperature, magnetic field) characteristic of the ISM. Consequently, the specific impact of Galactic morphology, such as the transition from the diffuse warm ionized medium (WIM) to dense supernova shells, on the formation of higher-order dressed solitons remains unexplored. Bridging this gap is essential for understanding how nonlinear wave structures respond to strong spatial gradients and inhomogeneities that are ubiquitous in realistic astrophysical plasmas.

In this paper, we extend the spatially dependent framework of Singh2026 to the higher-order nonlinear regime. We derive an extended KdV-type equation that incorporates both higher-order nonlinearity and dispersion, coupled with the realistic ISM parameter profiles developed in Singh2026. This approach allows us to: (1) obtain exact solutions for higher-order and dressed KA solitons in a structured plasma; (2) identify the specific Galactic regions where higher-order corrections dominate over first-order descriptions; and (3) analyze how suprathermal particle populations modify the dressing of solitons in different astrophysical environments.

The paper is organized as follows. Sec.~\ref{sec:model} outlines the structured ISM model. Sec.~\ref{subsec:fluid_model} discuss the basic fluid model. Sec.~\ref{sec:KdV} derives the leading order KdV equation  with reductive perturbation method. Sec.~\ref{sec:HO} discuss  the higher-order perturbation scheme, presents the derivation of the evolution equation, details the analytical solutions for dressed solitons. Sec.~\ref{sec:results} discusses the parametric analysis across different Galactic environments, and Sec.~\ref{Summary} summarizes our main conclusions.

\section{ISM Model Description}
\label{sec:model}

To connect the KA soliton properties to ISM environments, we employ the spatially structured ISM model introduced in Singh2026 and summarise it here for completeness. The model describes the large‑scale WIM, together with localised H\,{\sc ii} regions, SWBs, and SNRs. All equilibrium plasma parameters vary with Galactic cylindrical coordinates \((R,Z)\).

Any equilibrium parameter \(Q_0(R,Z)\) (density, magnetic field, temperature) is written as
\begin{equation}
Q_{\mathrm{0}}(R,Z)\equiv Q_{\mathrm{total}}(R,Z) = Q_{\mathrm{bkg}}(R,Z)
  + \sum_{i} Q_{\mathrm{struct},i}(R,Z),
\label{eq:super}
\end{equation}
where \(Q_{\mathrm{bkg}}\) is the smooth large‑scale WIM background and the sum runs over the embedded structures. The spatial variation of each structure is described by three analytical functions
\begin{align}
\mathcal{G}(d;A,\sigma) &= A\exp\!\left(-\frac{d^{2}}{2\sigma^{2}}\right), \\
\mathcal{P}(d;A,R_{\mathrm{flat}},s) &= A\exp\!\left(-\bigl(\tfrac{d}{R_{\mathrm{flat}}}\bigr)^{s}\right), \\
\mathcal{S}(d;A,R_{\mathrm{sh}},\sigma_{\mathrm{in}},\sigma_{\mathrm{out}}) &=
A\left\{\begin{array}{ll}
\exp\!\left(-\frac{(R_{\mathrm{sh}}-d)^{2}}{2\sigma_{\mathrm{in}}^{2}}\right), & d\le R_{\mathrm{sh}}\\[4pt]
\exp\!\left(-\frac{(d-R_{\mathrm{sh}})^{2}}{2\sigma_{\mathrm{out}}^{2}}\right), & d>R_{\mathrm{sh}}
\end{array}\right.
\end{align}
represent Gaussian, super‑Gaussian (plateau), and asymmetric shell profiles, respectively, and \(d_i = \sqrt{(R-R_i)^2+(Z-Z_i)^2}\) is the distance from the centre ($R_i, Z_i$) of structure \(i\). Numerical values of all amplitudes and scale parameters are listed in Table~\ref{tab:profiles_parameters}  (Appendix).\\
\noindent
\textbf{Background WIM profiles:}
The background electron density follows the NE2001 model \citep{cordes2002ne2001}:
\begin{equation}
n_{e,\mathrm{bkg}}(R,Z) = n_{1}\,e^{-|Z|/h_{1}} + n_{2}\,e^{-|Z|/h_{2} - (R-R_{\odot})/L_{2}},
\label{eq:nbkg}
\end{equation}
with $n_1 = 0.03\ \mathrm{cm}^{-3}$, $h_1 = 0.1\ \mathrm{kpc}$, $n_2 = 0.1\ \mathrm{cm}^{-3}$, $h_2 = 1.0\ \mathrm{kpc}$, $L_2 = 15.0\ \mathrm{kpc}$, $R_\odot = 8.5\ \mathrm{kpc}$. The background magnetic field follows the JF12 model \citep{Jansson2012}:
\begin{equation}
B_{0,\mathrm{bkg}}(R,Z) = B_{\odot}\,e^{-(R-R_{\odot})/L_{B}}\,e^{-|Z|/h_{B,1}}
   + \frac{B_{\mathrm{halo}}}{1+(|Z|/h_{B,2})^{2}},
\label{eq:Bbkg}
\end{equation}
with $B_\odot = 5.0\ \mu\mathrm{G}$, $L_B = 10.0\ \mathrm{kpc}$, $h_{B,1} = 1.0\ \mathrm{kpc}$, $B_{\text{halo}} = 2.0\ \mu\mathrm{G}$, $h_{B,2} = 3.0\ \mathrm{kpc}$. The background electron temperature is constant,
$T_{e,\mathrm{bkg}} = 8000\ \mathrm{K},$
representative of the WIM. The large‑scale positron fraction is
\begin{equation}
p_{\mathrm{large-scale}}(R) = p_{\mathrm{bkg}} + (p_{\mathrm{peak}}-p_{\mathrm{bkg}})\,\exp\!\left(-\frac{R^{2}}{2\sigma_{p,\mathrm{scale}}^{2}}\right),
\label{eq:pbkg}
\end{equation}
where $p_{\text{bkg}} = 0.01$, $p_{\text{peak}} = 0.2$, $\sigma_{p,\text{scale}} = 1.0\ \mathrm{kpc}$.

\noindent
\textbf{Total positron fraction:}
The total positron fraction is defined as $p_{\rm total}(R,Z)\equiv n_{\rm p,total}(R,Z)/n_{\rm i,total}(R,Z)$, where $n_{\rm p,total}(R,Z)$ and $n_{\rm i,total}(R,Z)$ are the local positron and ion densities respectively. Positrons are supplied by the large‑scale Galactic component and by the pulsar wind nebula (PWN) of the SNR. H\,{\sc ii} regions and SWBs do not contribute to pair production. Thus $p_{\mathrm{total}}(R,Z)$ is given as
\begin{align}
p_{\mathrm{total}}(R,Z) &= p_{\mathrm{large-scale}}(R) + p_{\mathrm{SNR}}(R,Z), \label{eq:p_total}\\
p_{\mathrm{SNR}}(R,Z) &= \mathcal{G}\bigl(d_{\mathrm{SNR}};A_{p,\mathrm{SNR}},\sigma_{p,\mathrm{SNR}}\bigr),\nonumber
\end{align}
with \(A_{p,\mathrm{SNR}}\) and \(\sigma_{p,\mathrm{SNR}}\) given in Table~\ref{tab:profiles_parameters}.

\noindent
\textbf{Total electron density:} The total electron density is the sum of the background WIM and the contributions from the H\,{\sc ii} region, SWB, and SNR:
\begin{align}
n_{e,\mathrm{total}} &=\; n_{e,\mathrm{bkg}}(R,Z) \nonumber\\ 
&\quad +\mathcal{G}\bigl(d_{\mathrm{HII}};A_{n,\mathrm{HII}},\sigma_{n,\mathrm{HII}}\bigr) \nonumber\\ \nonumber
&\quad +\underbrace{\mathcal{G}\bigl(d_{\mathrm{SWB}};A_{n,\mathrm{cav,SWB}},\sigma_{n,\mathrm{cav,SWB}}\bigr)}_{\text{cavity}}\\ 
&+\mathcal{S}\bigl(d_{\mathrm{SWB}};A_{n,\mathrm{sh,SWB}},R_{\mathrm{sh,SWB}},\sigma_{n,\mathrm{in,SWB}},\sigma_{n,\mathrm{out,SWB}}\bigr) \nonumber\\ \nonumber
&\quad +\underbrace{\mathcal{G}\bigl(d_{\mathrm{SNR}};A_{n,\mathrm{PWN,SNR}},\sigma_{n,\mathrm{PWN,SNR}}\bigr)}_{\text{PWN core}}\\
&+\mathcal{G}\bigl(d_{\mathrm{SNR}};A_{n,\mathrm{cav,SNR}},\sigma_{n,\mathrm{cav,SNR}}\bigr) \nonumber\\
&\quad +\mathcal{S}\bigl(d_{\mathrm{SNR}};A_{n,\mathrm{sh,SNR}},R_{\mathrm{sh,SNR}},\sigma_{n,\mathrm{in,SNR}},\sigma_{n,\mathrm{out,SNR}}\bigr).
\label{eq:ne_total}
\end{align}
The central PWN core introduces a strong magnetic spike, while the shell provides a compressed magnetic layer from the forward shock.

\noindent
\textbf{Total magnetic field:} The total magnetic field is constructed analogously, with a diamagnetic depression in the H\,{\sc ii} region and shell‑compressed fields in the SWB and SNR:
\begin{align}
B_{0,\mathrm{total}} &=\; B_{0,\mathrm{bkg}}(R,Z) \nonumber\\ \nonumber
&\quad +\mathcal{G}\bigl(d_{\mathrm{HII}};A_{B,\mathrm{HII}},\sigma_{B,\mathrm{HII}}\bigr)\\ \nonumber
&\quad +\underbrace{\mathcal{G}\bigl(d_{\mathrm{SWB}};A_{B,\mathrm{cav,SWB}},\sigma_{B,\mathrm{cav,SWB}}\bigr)}_{\text{cavity}}\\ \nonumber
&+\mathcal{S}\bigl(d_{\mathrm{SWB}};A_{B,\mathrm{sh,SWB}},R_{\mathrm{sh,SWB}},\sigma_{B,\mathrm{in,SWB}},\sigma_{B,\mathrm{out,SWB}}\bigr)\\ \nonumber
&\quad +\underbrace{\mathcal{G}\bigl(d_{\mathrm{SNR}};A_{B,\mathrm{PWN,SNR}},\sigma_{B,\mathrm{PWN,SNR}}\bigr)}_{\text{PWN core}}\\ 
&+\mathcal{S}\bigl(d_{\mathrm{SNR}};A_{B,\mathrm{sh,SNR}},R_{\mathrm{sh,SNR}},\sigma_{B,\mathrm{in,SNR}},\sigma_{B,\mathrm{out,SNR}}\bigr).
\label{eq:B_total}
\end{align}
The central PWN core introduces a strong magnetic spike, while the shell provides twin peaks from shock compression.

\noindent
\textbf{Total electron temperature:} The temperature combines the cool WIM background with hot, shock‑heated gas in the H\,{\sc ii} region, the flat‑top plateau of the SWB interior, and the multi‑component thermal structure of the SNR:
\begin{align}
T_{e,\mathrm{total}} &=\; T_{e,\mathrm{bkg}} \nonumber\\ \nonumber
&\quad +\mathcal{G}\bigl(d_{\mathrm{HII}};A_{T,\mathrm{HII}},\sigma_{T,\mathrm{HII}}\bigr) \nonumber\\\nonumber
&\quad +\underbrace{\mathcal{P}\bigl(d_{\mathrm{SWB}};A_{T,\mathrm{plat,SWB}},R_{\mathrm{flat,SWB}},s=8\bigr)}_{\text{hot plateau}}\\
&+\mathcal{S}\bigl(d_{\mathrm{SWB}};A_{T,\mathrm{sh,SWB}},R_{\mathrm{sh,SWB}},\sigma_{T,\mathrm{in,SWB}},\sigma_{T,\mathrm{out,SWB}}\bigr) \nonumber\\ \nonumber
&\quad +\underbrace{\mathcal{G}\bigl(d_{\mathrm{SNR}};A_{T,\mathrm{core,SNR}},\sigma_{T,\mathrm{core,SNR}}\bigr)}_{\text{hot interior}}\\
&+\mathcal{S}\bigl(d_{\mathrm{SNR}};A_{T,\mathrm{sh,SNR}},R_{\mathrm{sh,SNR}},\sigma_{T,\mathrm{in,SNR}},\sigma_{T,\mathrm{out,SNR}}\bigr).
\label{eq:Te_total}
\end{align}
The temperature rises sharply in the SNR shell and remains elevated in the cavity due to the Sedov‑Taylor expansion \citep{CHEVALIERInteraction1977}.

\noindent
\textbf{Derived equilibrium quantities:}
For notational brevity, we define the total equilibrium profiles as
\begin{align}
n_{e0} &\equiv n_{e,\mathrm{total}}(R,Z), 
&\quad 
B_{0} &\equiv B_{0,\mathrm{total}}(R,Z), \nonumber \\  \nonumber 
T_{e0} &\equiv T_{e,\mathrm{total}}(R,Z), 
&\quad
p_{0} &\equiv p_{\mathrm{total}}(R,Z).
\end{align}
The charge neutrality then gives the equilibrium ion density
\begin{equation}
n_{i0}= \frac{n_{e0}}{1+p_{0}}.
\label{eq:iondens}
\end{equation}
Outside SNR cores \(p_{0}\) is negligible and \(n_{i0}\approx n_{e0}\); inside composite SNRs,  the PWN can make $p_{0}\gg1$, leading to ion depletion. The local Alfvén speed is \(V_A = B_{0}/\sqrt{4\pi n_{i0}m_i}\), which enters the dispersive coefficient of the KdV equation and the higher‑order coefficients through normalization.

The local plasma beta is written as
\begin{equation}
\beta = \frac{8\pi n_{i0}\,k_B T_{e0}}{B_{0}^{2}},
\label{eq:beta}
\end{equation}
which can be evaluated at each point in \((R,Z)\) map. In the present low‑\(\beta\) KAW framework, we restrict our attention to the regime ${m_e}/{m_i} < \beta < 1$,
where the KA mode and its solitary solutions are admissible. Regions where \(\beta > 1\) (the thermally over‑pressured interiors of the H\,{\sc ii} region, SWB, and SNR) form exclusion zones (EZs) in which no admissible KA soliton exists. The spatial distribution of these EZs and the morphological variety of solitons in the admissible regions are discussed in Sec.~\ref{sec:results}.

\section{Fluid model and physical assumptions}
\label{subsec:fluid_model}

To describe the local propagation of KA waves, we introduce a Cartesian coordinate system $(x,z)$ attached to each Galactic position $(R,Z)$. Over the microscopic scale of the soliton, the ambient Galactic magnetic field is approximated as uniform and taken to define the local $z$-direction. This approximation is justified by the large separation between the characteristic kinetic-Alfv\'en scale, of order the ion gyroradius ($\rho_i \sim 200$~km), and the kiloparsec-scale variation of the background ISM structure \citep{SINGHInterstellar2026}. The plasma is modeled locally as a collisionless, magnetized electron--positron--ion (e-p-i) system. The ion component is treated as a cold fluid, consistent with the low-$\beta$ WIM regime in which magnetic pressure dominates over ion thermal pressure \citep{SINGHInterstellar2026}. Electrons and positrons are assumed inertialess and described by superthermal $\kappa$ distributions, allowing for non-Maxwellian high-energy tails generated by shocks, turbulent energization, and related processes in the structured ISM \citep{VASYLIUNASSurvey1968,DRURYIntroduction1983}.

All coefficients entering the reduced wave description is evaluated from the local plasma state supplied by the ISM model.
Consequently, the nonlinear and dispersive coefficients of the evolution equation become explicit functions of Galactic location, so that the polarity, amplitude, width, and morphology of the soliton are all controlled by the surrounding ISM environment. 
The governing fluid equations used for this analysis are introduced below in this section. In the following section, these equations are treated with the reductive perturbation method (RPM) to derive the nonlinear evolution equation and inhomogeneous KdV equation.
Unlike \citet{SINGHInterstellar2026}, where the perturbation expansion was truncated at leading order and yielded a standard KdV equation, we retain the next order in the smallness parameter $\epsilon$. This extension captures the higher-order nonlinear and dispersive contributions responsible for dressing the soliton core. The resulting inhomogeneous KdV-type equation supports closed-form dressed soliton solutions whose morphology is substantially richer than the symmetric, sign-definite profiles of the leading-order theory.

To describe the low-$\beta$ KA perturbations within this present fluid description, we adopt the standard two-potential formulation \citep{Kadomtsev1965}, in which $\psi$ and $\phi$ denote the parallel and perpendicular electrostatic potentials, respectively. The normalized continuity and momentum equations \citep{SINGHInterstellar2026} governing the cold ion fluid are
\begin{equation}
\frac{\partial n_{i}}{\partial t}
  +\frac{\partial(n_{i}v_{ix})}{\partial x}
  +\frac{\partial(n_{i}v_{iz})}{\partial z}=0,
\label{eq1}
\end{equation}
\begin{equation}
\frac{\partial v_{iz}}{\partial t}
  +v_{ix}\frac{\partial v_{iz}}{\partial x}
  +v_{iz}\frac{\partial v_{iz}}{\partial z}
  =-{\Lambda}\frac{\partial\psi}{\partial z},
\label{eq2}
\end{equation}
\begin{equation}
v_{ix}=-{\Lambda}\frac{\partial^{2}\phi}{\partial x\,\partial t},
\label{eq3}
\end{equation}
\begin{equation}
{\Lambda}\frac{\partial^{4}(\phi-\psi)}{\partial x^{2}\partial z^{2}}
  =\frac{\partial^{2}n_{i}}{\partial t^{2}}
  +\frac{\partial^{2}(n_{i}v_{iz})}{\partial z\,\partial t},
\label{eq4}
\end{equation}
where ${\Lambda}=\beta/2$. The normalizations are: number densities $n_j=n_j^{\prime}/n_{j0}$ ($j=i,e,p$); wave potentials $(\psi,\phi)=e(\psi^{\prime},\phi^{\prime})/(k_BT_{e0})$;
time $t=t^{\prime}\Omega_i$; spatial coordinates
$(x,z)=(x^{\prime},z^{\prime})\Omega_i/V_A$; and fluid velocity
$v_{i(x,z)}=v_{i(x,z)}^{\prime}/V_A$, where 
$\Omega_i=eB_0/(m_ic)$ is the local ion cyclotron frequency. The primed quantities denote the corresponding dimensional physical variables prior to normalization.

The equilibrium charge neutrality condition can be written as
\begin{equation}
\delta_{ei}=1+p_0,
\label{eq5}
\end{equation}
where $\delta_{ei}\equiv n_{e0}/n_{i0}$ is the equilibrium electron-to-ion density ratio. This relation expresses local quasineutrality in the e-p-i plasma: the positive ion charge is balanced by the combined contribution of electrons and positrons, so that any finite positron content necessarily increases the required electron density relative to the ion density. In the limit $p_0=0$, the standard electron--ion neutrality condition $\delta_{ei}=1$ is recovered. In the present ISM model, $p_0$ is negligible throughout most of the WIM, H\,\textsc{ii} regions, and SWB environments, so the plasma behaves effectively as an ordinary electron--ion system. However, inside composite SNR interiors containing a pulsar wind nebula, relativistic pair injection can drive $p_0$ to appreciable values, leading to $\delta_{ei}>1$ and substantial ion dilution. This modification is physically important because it changes the local inertia carried by the ion component and therefore affects the Alfv\'en speed, the dispersive scale, and ultimately the nonlinear coefficients governing KA soliton evolution.

The electron and positron populations are assumed to obey superthermal $\kappa$ distributions \citep{VASYLIUNASSurvey1968}. Their corresponding normalized number densities can then be expanded as power series in the electrostatic potential $\psi$, giving
\begin{equation}
n_{e}=\!\left[1-\frac{\psi}{\kappa_{e}-\tfrac{3}{2}}
  \right]^{-\kappa_{e}+\frac{1}{2}}
  \approx 1+c_{1}\psi+c_{2}\psi^{2}+c_3\psi^3+{\cal O}(\psi^4),
\label{eq6}
\end{equation}
and
\begin{equation}
n_{p}=\!\left[1+\frac{\sigma\psi}{\kappa_{p}-\tfrac{3}{2}}
  \right]^{-\kappa_{p}+\frac{1}{2}}
  \approx 1-d_{1}\psi+d_{2}\psi^{2}-d_3\psi^3+{\cal O}(\psi^4),
\label{eq7}
\end{equation}
where 
\begin{equation}
c_{1}=\frac{K_{e1}}{K_{e3}},\quad c_{2}=\frac{K_{e1} K_{e2}}{2K_{e3}^{2}}, \quad c_3 = \frac{K_{e1}K_{e2}K_{e4}}{6 K_{e3}^3}, \nonumber
\end{equation}
with
\begin{equation*}
\begin{aligned}
K_{e1}&\equiv(\kappa_{e}-\tfrac{1}{2}), & K_{e2}&\equiv(\kappa_{e}+\tfrac{1}{2}), \\
K_{e3}&\equiv(\kappa_{e}-\tfrac{3}{2}), & K_{e4}&\equiv(\kappa_{e}+\tfrac{3}{2});
\end{aligned}
\end{equation*}
\begin{equation}
d_{1}=\sigma\frac{K_{p1}}{K_{p3}},\quad d_{2}=\sigma^2\frac{K_{p1} K_{p2}}{2K_{p3}^{2}}, \quad d_3 = \sigma^3\frac{K_{p1}K_{p2}K_{p4}}{6 K_{p3}^3}, \nonumber
\end{equation}
with
\begin{equation*}
\begin{aligned}
K_{p1}&\equiv(\kappa_{p}-\tfrac{1}{2}), & K_{p2}&\equiv(\kappa_{p}+\tfrac{1}{2}), \\
K_{p3}&\equiv(\kappa_{p}-\tfrac{3}{2}), & K_{p4}&\equiv(\kappa_{p}+\tfrac{3}{2}),
\end{aligned}
\end{equation*}

\noindent
and $\sigma =T_{e0}/T_{p0}$ is the electron-to-positron temperature ratio. Here $\kappa_e$ and $\kappa_p$ are the spectral indices for the electron and positron $\kappa$ distributions, respectively; smaller values of $\kappa$ correspond to harder suprathermal tails, while $\kappa\to\infty$ recovers the Maxwellian limit. In the present analysis, $\kappa_e$ is the primary control parameter of interest, as it governs the electron kinetic response that determines both the nonlinear steepening coefficient and the character of the higher-order correction terms.

\section{Leading-Order KdV Equation}
\label{sec:KdV}

To extract a nonlinear evolution equation from the fluid system
(\ref{eq1})--(\ref{eq7}), we employ the reductive perturbation method
\citep{TANIUTIReductive1968,TANIUTIPerturbation1969}.  The independent variables are stretched into the slow, 
co-moving frame via
\begin{equation}
\xi=\epsilon^{1/2}(l_{x}x+l_{z}z-\lambda t), \quad \tau=\epsilon^{3/2}\,t,
\label{eq8}
\end{equation}

where $\lambda$ is the phase velocity of the KA mode, $\epsilon\ll1$ is a small dimensionless parameter that measures the weakness of nonlinearity and dispersion, and $l_x$, $l_z$ are the direction cosines of the propagation vector in the $x$--$z$ plane, with $l_x^2+l_z^2=1$.  The oblique propagation geometry ($l_x\neq0$, $l_z\neq0$) is essential for KA dynamics, as the finite perpendicular component $k_\perp=l_x k$ provides the dispersive mechanism via the finite Larmor radius effect.

The fluid variables are expanded about their equilibrium values as power series in $\epsilon$:

\begin{align}
v_{ix}&=\epsilon v_{ix}^{(1)}+\epsilon^{2}v_{ix}^{(2)}+\cdots,
&v_{iz}&=\epsilon v_{iz}^{(1)}+\epsilon^{2}v_{iz}^{(2)}+\cdots, \label{eq10}  \\
\psi&=\epsilon\psi^{(1)}+\epsilon^{2}\psi^{(2)}+\cdots,
&\phi&=\phi^{(1)}+\epsilon\phi^{(2)}+\cdots.
\label{eq11}
\end{align}

Using the plasma charge-neutrality approximation and the $\kappa$-distribution expansions (\ref{eq6})--(\ref{eq7}), the ion number density is
\begin{equation}
n_{i}=(1+p_0)\,n_e-p_0\,n_p
     =1+a_{1}\psi+a_{2}\psi^{2}+\cdots,
\label{eq12}
\end{equation}
where
\begin{equation*}
\begin{aligned}
a_1 &= (1+p_0)c_1 + p_0 d_1, \quad & a_2 &= (1+p_0)c_2 - p_0 d_2, \\
a_3 &= (1+p_0)c_3 + p_0 d_3.
\end{aligned}
\end{equation*}
\noindent
Since $a_1-a_3$ depend on $p_0$, and the $\kappa$ indices, they vary with Galactic position through the structured
ISM model.

Substituting the stretched variables (\ref{eq8})--(\ref{eq12}) into the fluid equations (\ref{eq1})--(\ref{eq4}) and collecting terms at $\mathcal{O}(\epsilon^{3/2})$ yields four coupled first-order perturbation equations for $v_{ix}^{(1)}$, $v_{iz}^{(1)}$, $\psi^{(1)}$, and $\phi^{(1)}$ (listed in full in Appendix~\ref{app:FO}). Eliminating the perturbed fields from those equations yields the biquadratic
dispersion relation
\begin{equation}
a_{1}\lambda^{4}-(a_{1}+{\Lambda})l_{z}^{2}\lambda^{2}+{\Lambda} l_{z}^{4}=0.
\label{eq14}
\end{equation}
Equation~(\ref{eq14}) admits two physically distinct roots.
The first corresponds to the KA mode:
\begin{equation}
\lambda^{2}=l_{z}^{2},
\label{eq15}
\end{equation}
and the second to the ion-acoustic mode:
\begin{equation}
\lambda^{2}=\frac{{\Lambda}\,l_{z}^{2}}{a_{1}}.
\label{eq16}
\end{equation}
In the dispersionless limit, the KA phase velocity (Eq.~\ref{eq15}) is independent of the suprathermality indices and particle composition, while the ion-acoustic phase velocity (Eq.~\ref{eq16}) depends on both through $a_1$. We select the KA branch throughout, as it governs the obliquely propagating dispersive Alfv\'enic perturbations of interest.  The first-order perturbed fields obtained from Appendix~\ref{app:FO} using $\lambda^2=l_z^2$ serve as forcing terms in the second-order system of Appendix~\ref{app:SO}.

Proceeding to $\mathcal{O}(\epsilon^{5/2})$ in Eqs.~(\ref{eq1})--(\ref{eq4}), we obtain the second-order perturbation equations, which are listed explicitly in Appendix~\ref{app:SO}. Eliminating the second-order quantities $v_{ix}^{(2)}$, $v_{iz}^{(2)}$, $\psi^{(2)}$, and $\phi^{(2)}$ with the help of first-order solutions given in Appendix~\ref{app:FO}$,$ we arrive at the KdV equation governing the leading-order potential $\psi^{(1)}$:
\begin{equation}
\frac{\partial\psi^{(1)}}{\partial\tau}
  +{\cal A}\,\psi^{(1)}\frac{\partial\psi^{(1)}}{\partial\xi}
  +{\cal B}\,\frac{\partial^{3}\psi^{(1)}}{\partial\xi^{3}}=0.
\label{eq17}
\end{equation}
The nonlinear coefficient ${\cal A}$ and the dispersion coefficient ${\cal B}$ are
\begin{equation}
{\cal A}(R,Z)=-a_{1}l_{z},
\label{eq18}
\end{equation}
\begin{equation}
{\cal B}(R,Z)=-\frac{l_{x}^{2}l_{z}{\Lambda}}{2(a_{1}-{\Lambda})}.
\label{eq19}
\end{equation}
Because $a_1$ and $\Lambda$ inherit the full spatial dependence of the ISM model through Sec.~\ref{sec:model}, the coefficients ${\cal A}$ and ${\cal B}$ are themselves functions of Galactic position.

The stationary localized solution of Eq.~(\ref{eq17}) is obtained by applying the traveling-wave ansatz $\eta=\xi-U\tau$ and integrating twice with decaying boundary conditions: 
\begin{equation}
\psi^{(1)}=\psi_{0}\,\mathrm{sech}^{2}\!\left(\frac{\eta}{W}\right),
\label{eq20}
\end{equation}
where $\psi_{0}={3U}/{{\cal A}}$ is the peak amplitude, $W=\sqrt{{4{\cal B}}/{U}}$ is the soliton width and $U$ is the soliton frame speed. Since ${\cal A}$ and ${\cal B}$ vary with Galactic position, both $\psi_0$ and $W$ are intrinsic functions of location.

The symmetric $\mathrm{sech}^{2}$ profile of Eq.~(\ref{eq20}) is a fundamental limitation of the first-order KdV description: it cannot produce double-humped, asymmetric, or mixed-polarity profiles regardless of the plasma parameters. In energetic ISM environments, such as the inner Galactic region or the vicinity of SWB and SNR shells, where the second-order correction $\psi^{(2)}$ is comparable in magnitude to $\psi^{(1)}$, the full dressed-soliton profile must be computed from the higher-order equation derived in the following section.

\section{Higher-Order Dressed KA Soliton}
\label{sec:HO}

\subsection{KdV-type inhomogeneous equation}
\label{subsec:iKdV}

To account for corrections beyond the leading KdV balance, the perturbation expansion is extended to $\mathcal{O}(\epsilon^{7/2})$.  Collecting terms of this order in Eqs.~(\ref{eq1})--(\ref{eq4}) yields four equations coupling the third-order quantities $v_{ix}^{(3)}$, $v_{iz}^{(3)}$, $\psi^{(3)}$, and $\phi^{(3)}$ to nonlinear products of first- and second-order equations; these equations are listed in Appendix~\ref{app:TO}. Eliminating the third-order perturbed quantities from these equations by employing the first-order solutions from Appendix~\ref{app:FO} and the second-order solutions from Appendix~\ref{app:SO}, and performing the standard algebraic manipulation \citep{Kodama1978,Sugimoto1977}, yields the following inhomogeneous KdV-type equation governing the second-order correction $\psi^{(2)}$:
\begin{equation}
\frac{\partial\psi^{(2)}}{\partial\tau}
  +{\cal A}\frac{\partial\!\left(\psi^{(1)}\psi^{(2)}\right)}{\partial\xi}
  +{\cal B}\frac{\partial^{3}\psi^{(2)}}{\partial\xi^{3}}
  =G\!\left(\psi^{(1)}\right).
\label{eq23}
\end{equation}
The left-hand side has the same operator structure as the linearization of the KdV equation about $\psi^{(1)}$; the inhomogeneous right-hand side is entirely determined by the leading-order soliton and takes the form
\begin{eqnarray}
G(\psi^{(1)})
  = A_{1}\frac{\partial{\left(\psi^{(1)}\right)}^{3}}{\partial\xi}
  + A_{2}\frac{\partial}{\partial\xi}\!\left(
    \psi^{(1)}\frac{\partial^{2}\psi^{(1)}}{\partial\xi^{2}}\right)
\nonumber\\
  + A_{3}\frac{\partial}{\partial\xi}\!\left(
      \frac{\partial\psi^{(1)}}{\partial\xi}\right)^{\!2}
  - A_{4}\frac{\partial^{5}\psi^{(1)}}{\partial\xi^{5}}.
\label{G_psi}
\end{eqnarray}
The coefficients $A_1$--$A_4$ depend on the leading-order KdV coefficients ${\cal A}$ and ${\cal B}$ and, the equilibrium plasma parameters; their explicit expressions are given in Appendix~\ref{app:coeffs}. 
The first term represents the cubic nonlinearity: it is the spatial derivative of the amplitude cubed, the leading higher-order nonlinear contribution arising at $\mathcal{O}(\epsilon^{7/2})$.  The second term is a nonlinear-dispersive cross-term that couples the wave amplitude to its own curvature. The third term is a gradient-squared nonlinearity that couples the local slope to the local curvature of the wave profile.  All three of these terms are of the same perturbative order $\mathcal{O}(\epsilon^{7/2})$; they differ in the spatial structure, amplitude, amplitude--curvature, or gradient--curvature, through which nonlinearity enters at this order.  The fourth term, $-A_4\,\partial^5\psi^{(1)}/\partial\xi^5$, is qualitatively different: it is linear in $\psi^{(1)}$ and represents fifth-order linear dispersion \citep{KAWAHARAOscillatory1972}, which becomes the dominant dispersive mechanism at this order once the standard second-order (KdV) balance has been satisfied at leading order. It is the interplay between the three nonlinear contributions and this fifth-order dispersive term that allows $\psi^{(2)}$ to develop off-center extrema of opposite polarity to the leading-order core, producing the composite dressed morphologies
mapped in Sec.~\ref{sec:results}.

\subsection{Dressed soliton solution}
\label{subsec:dressed_sol}

The total electrostatic potential is the superposition of the leading-order ($\psi^{(1)}$) and higher‑order contribution ($\psi^{(2)}$):
\begin{equation}
\psi_{\mathrm{Total}}=\psi^{(1)}+\psi^{(2)}.
\label{eq24}
\end{equation}
Substituting the KdV soliton solution~(\ref{eq20}) as a particular solution of the homogeneous part of Eq.~(\ref{eq23}) and solving for $\psi^{(2)}$ by the method of \citet{She07} yields the closed-form dressed soliton:
\begin{eqnarray}
\psi_{\mathrm{Total}}
  &=& \psi_{0}\,\mathrm{sech}^{2}\!\left(\frac{\eta}{W}\right)
+\;\frac{\psi_{0}^{2}}{UD^{2}}\,\alpha_{11}\,
     \mathrm{sech}^{2}\!\left(\frac{\eta}{D}\right)
     \nonumber \\
   &&-\alpha_{2}\,\frac{\psi_{0}^{2}}{UD^{2}}\,
     \mathrm{sech}^{4}\!\left(\frac{\eta}{D}\right),
\label{eq25}
\end{eqnarray}
where the composite coefficients are
\begin{equation}
\alpha_{2}
  =\frac{3{\cal B}}{{\cal A}}A_{1}-3A_{2}-2A_{3}-\frac{10{\cal A}}{{\cal B}}A_{4},
\label{eq26}
\end{equation}
\begin{equation}
\alpha_{10}
  =\frac{3{\cal B}A_{1}}{{\cal A}}-\frac{1}{2}A_{2}-\frac{2}{3}A_{3}
   -\frac{10{\cal A}}{3{\cal B}}A_{4},
\label{eq27}
\end{equation}
\begin{equation}
\alpha_{11}=\alpha_{10}+\alpha_{2}.
\label{eq28}
\end{equation}
The soliton speed $U$ is related to the Mach-number $M$ by the relation $U+\Delta U= M - 1=\displaystyle\Delta M$
\begin{equation}
U=\frac{{\cal B}^{2}}{8A_{4}}\!\left[\!\left(1+\frac{16A_{4}\,\Delta M}{{\cal B}^{2}}
  \right)^{\!1/2}-1\right],
\label{eq29}
\end{equation}
and the higher-order soliton width is expressed as
\begin{equation}
D^{-1}=\sqrt{\frac{\Delta M}{2{\cal B}}}\;\frac{{{\cal B}^{2}}}{\displaystyle
  {{\cal B}^{2}}+{2A_{4}\Delta M}}.
\label{HO_width}
\end{equation}

\noindent
Equation~(\ref{eq25}) gives the closed-form expression for the dressed higher-order soliton potential. It describes the total potential as a superposition of a $\mathrm{sech}^2$ core (amplitude $\psi_0$, width $W$, controlled by the leading-order KdV balance) and two correction terms with width $D$ whose signs and magnitudes are set by the spatially varying coefficients $\alpha_{11}$ and $\alpha_2$ through ${\cal A}$, ${\cal B}$, and $A_1$--$A_4$.  

\section{Results and Parametric Analysis}
\label{sec:results}
In this section, we analyze the morphological evolution and existence domains of higher order KA solitons across the Galactic plane by solving the higher-order KdV-type system derived in the preceding sections. The total electrostatic potential $\psi(\eta) \equiv \psi_\mathrm{Total}(\eta)$ depends explicitly on the spatially varying ISM parameters at each Galactic position ($R,Z$). The resulting composite solution combines sech$^2$ and sech$^4$ contributions, with relative weights set by the local ISM coefficients. This interplay yields a far richer catalog of morphologies than the symmetric, sign‑definite profiles permitted by leading‑order KdV theory alone.
\subsection{Classification Algorithm and Diagnostic Framework}
In order to map the full solution space systematically, we evaluate $\psi(\eta)$ on a uniform two-dimensional grid covering  $R \in [0,16]$~kpc and $Z \in [-2.5,2.5]$~kpc, using a baseline resolution  of $1000\times1000$ points in the $(R,Z)$ plane. This corresponds to spatial samplings of $\Delta R \simeq 4$~pc and $\Delta Z \simeq 16$~pc, which are sufficient to resolve the large-scale morphology of the structured ISM model and the associated soliton-class boundaries. Tests with finer grids show no significant change in the dominant morphology domains, indicating numerical convergence at the level required for the present analysis.

At each grid point, the local ISM model supplies the equilibrium plasma parameters ($n_{i0}$, $T_{e0}$, $B_0$, and $\beta$), which determine the coefficients ${\cal A}$, ${\cal B}$, and $A_1$--$A_4$ of the higher-order evolution equations. Before computing a solution, two admissibility conditions are imposed. First, the physical constraint $m_e/m_i < \beta < 1$ must be satisfied. Second, the higher-order coefficients must remain regular and the soliton width $D$ must be finite and positive. Points failing either condition are excluded and marked as EZs in Fig.~\ref{fig:2d_morphology}.

At each admissible location, the full potential profile $\psi(\eta)$ is evaluated over a sufficiently large domain in $\eta$ such that $\psi(\eta)\to 0$ at the boundaries. To characterize the solution morphology, we extract three diagnostic quantities: the central value $\psi(0)$, the largest positive extremum $\psi_{\max} \equiv \max_{\eta} \psi(\eta)$, and the largest negative extremum $\psi_{\min} \equiv \min_{\eta} \psi(\eta)$.
Here, $\psi_{\max}$ and $\psi_{\min}$ are defined as the dominant extrema associated with the localized soliton structure, excluding the asymptotic background where $\psi(\eta) \rightarrow 0$. In particular, for single-hump solutions, $\psi_{\max}$ coincides with $\psi(0)$, while $\psi_{\min}$ remains close to zero and does not correspond to a physically meaningful extremum of the soliton itself.

The solutions are classified according to the morphology of $\psi(\eta)$ using the central value $\psi(0)$ and the dominant positive and negative extrema of the localized profile, as defined below:

\begin{itemize}
    \item $\psi_{\rm I}$ (positive single-hump): $\psi(0)>0$, with a single dominant maximum at $\eta=0$ such that $\psi(0)=\psi_{\max}$ and no secondary extrema exist.
    \item $\psi_{\rm II}$ (positive double-hump): $\psi(0)>0$, with two symmetric off-center maxima $\psi(\pm\eta_m)=\psi_{\max}$, while $\psi(0)$ is smaller than the peak amplitude.
    \item $\psi_{\rm III}$ (negative double-hump): $\psi(0)<0$, with two symmetric off-center minima $\psi(\pm\eta_m)=\psi_{\min}$, and the central value not coinciding with the dominant extrema.
    \item $\psi_{\rm IV}$ (dressed soliton): $\psi(0)>0$, with a central maximum $\psi(0)=\psi_{\max}$ accompanied by off-center minima $\psi_{\min}<0$, indicating the presence of opposite-polarity side lobes generated by higher-order corrections.
    \item $\psi_{\rm V}$ (split/degenerate double-hump): the dominant extrema occur at $\eta=\pm\eta_m$, while the central value $\psi(0)$ is significantly suppressed relative to $|\psi_{\max}|$ or $|\psi_{\min}|$, indicating a strong splitting of the soliton structure.
\end{itemize}
Figure~\ref{fig:1d_profiles} presents the representative potential profiles for each class $\psi_{\rm I-V}$, with the labels identifying the respective morphology.

No robust instances of negative single-hump solutions are found within the explored parameter space; therefore, this class is not included as a distinct category in the morphology maps. 
Profiles that do not satisfy any of the above criteria occur only near parameter boundaries where the solution becomes weakly localized or numerically sensitive. These cases are rare and do not form coherent spatial domains; they are therefore not treated as a separate physical class in the present analysis.
This classification procedure converts the full $(R,Z)$ parameter scan into the morphology map shown in Fig.~\ref{fig:2d_morphology}, where each color represents a distinct nonlinear solution class.

\subsection{Soliton Profiles at Representative Galactic Radii}

\begin{figure}[!htbp]
\centering
\includegraphics[width=0.47848465\textwidth,height=0.468\textwidth]{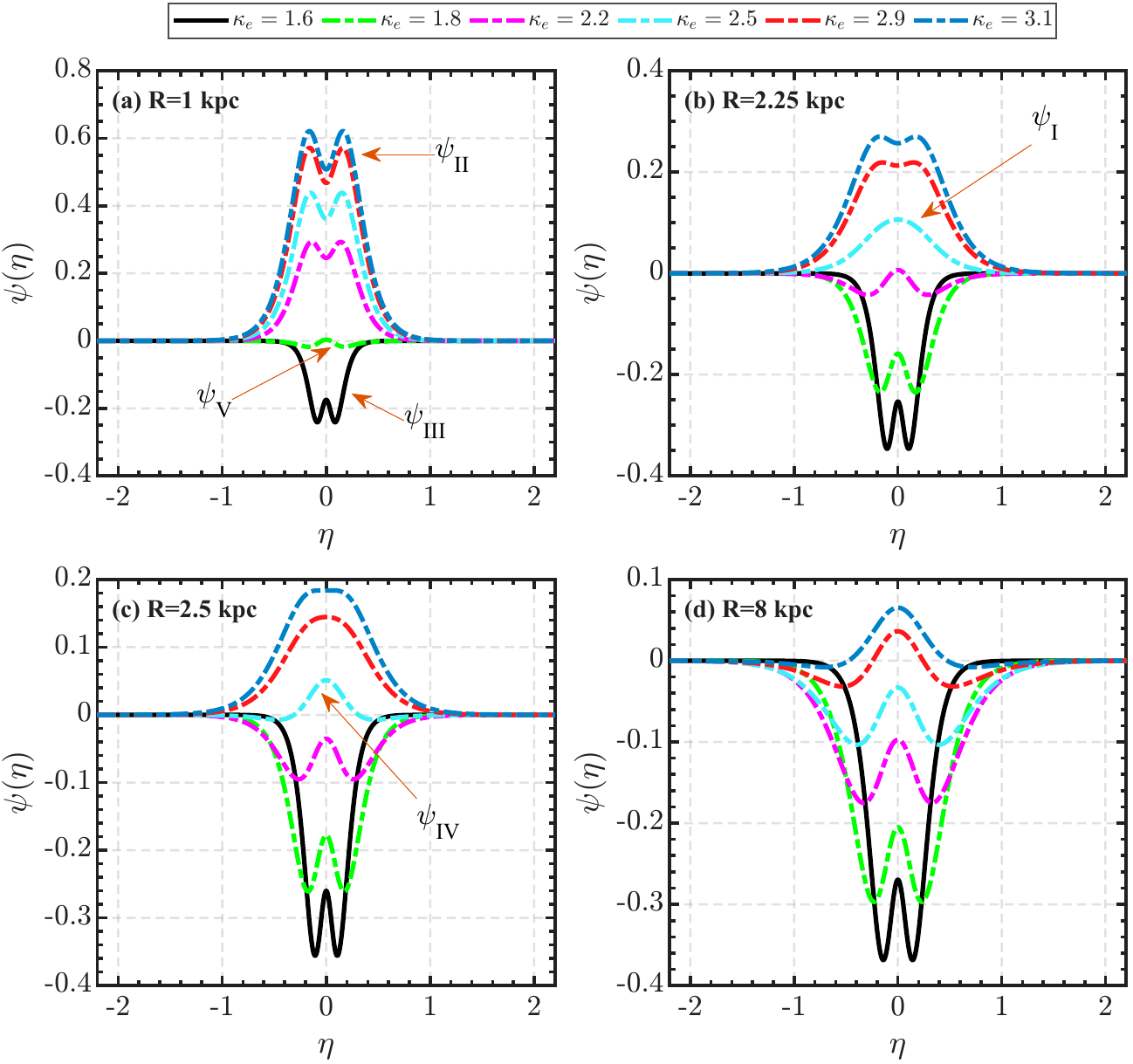}
\caption{Representative potential profiles $\psi(\eta)$ at selected Galactic locations, categorizing the structural transitions between positive single-hump ($\psi_{\rm I}$), positive double-hump ($\psi_{\rm II}$), negative double-hump ($\psi_{\rm III}$), dressed ($\psi_{\rm IV}$), and split-core ($\psi_{\rm V}$) states across varying $\kappa_e$, at fixed $\theta=80^\circ$, $M=-0.01$ and $\sigma=1$.}
\label{fig:1d_profiles}
\end{figure}

Figure~\ref{fig:1d_profiles} provides the profile-level confirmation of the morphological classification and the amplitude diagnostics. This figure resolves the full one-dimensional potential $\psi(\eta)$ across the co-moving coordinate $\eta$ for four sampled Galactic radii ($R = 1.0, 2.25, 2.5, 8.0$~kpc) and six spectral indices ($\kappa_e \in \{1.6, 1.8, 2.2, 2.5, 2.9, 3.1\}$). These locations are selected to sample the diverse plasma environments defined by our structured ISM model, illustrating the direct regulation of waveform architecture by local background parameters.  Within the inner Galactic region at $R = 1.0$~kpc (Fig.~\ref{fig:1d_profiles}a), the wave assumes a negative double-hump morphology ($\psi_{\rm III}$) at the strongly suprathermal limit of $\kappa_e = 1.6$.  
As $\kappa_e$ increases, the potential undergoes a polarity reversal into a positive double-hump state ($\psi_{\rm II}$) for $\kappa_e > 1.8$, passing through a split-core $\psi_{\rm V}$ phase marked by the green-dashed curve at $\kappa_e = 1.8$. 
At $R = 2.25$~kpc (Fig.~\ref{fig:1d_profiles}b), the structural identity of the soliton exhibits high sensitivity to the electron distribution.
The profiles maintain a $\psi_{\rm III}$ character at $\kappa_e = 1.6$ and 1.8, transitioning into a suppressed-core intermediate state ($\psi_{\rm V}$) at $\kappa_e = 2.2$. 
Notably, at $\kappa_e = 2.5$, the higher-order corrections balance such that the wave recovers the standard, single-peaked KdV-type profile ($\psi_{\rm I}$) predicted in first-order theory \citep{SINGHInterstellar2026}. 
Subsequent increases in $\kappa_e$ to 2.9 and 3.1 transform this single-peaked solution back into a $\psi_{\rm II}$ morphology, where the off-center maxima $\psi_{\max}$ are comparable to the core value $\psi(0)$. 
At $R = 2.5$~kpc (Fig.~\ref{fig:1d_profiles}c), the negative double-hump state $\psi_{\rm III}$ dominates the suprathermal regime for $\kappa_e \le 2.2$.  
At the intermediate value of $\kappa_e = 2.5$, the potential clearly reveals the dressed soliton morphology ($\psi_{\rm IV}$). 
Further thermalization leads to $\psi_{\rm I}$ type solutions, though the profile for $\kappa_e = 3.1$ (blue-dashed curve) exhibits a characteristic flattening of the peak compared to standard KdV pulses.
Sampling the diffuse outer disk at $R = 8.0$~kpc (Fig.~\ref{fig:1d_profiles}d), beyond the influence of the SWB shell, the waveforms deviate significantly from their inner-Galaxy counterparts.  
Here, the system supports $\psi_{\rm III}$ profiles for $\kappa_e \le 2.5$, transitioning into the dressed $\psi_{\rm IV}$ state at $\kappa_e \in \{2.9, 3.1\}$.  
These waveforms are slightly wider than those observed at smaller radii, a result of the expanded dispersive length scales in the lower-density  environments of the outer disk. 
The enhanced influence of higher-order dressing is evident across all $\kappa_e$ values, as the profiles remain non-standard and fail to recover the symmetric $\text{sech}^2$ geometry of the $\psi_{\rm I}$ class at this location. 
Having established the fundamental waveform classes and their dependence on local plasma conditions, we now project these morphological states across the full Galactic plane to resolve their global spatial distribution.

\subsection{Global Morphology Map}
\begin{figure}[!htbp]
\centering
\includegraphics[width=0.456838\textwidth,height=0.65\textwidth]{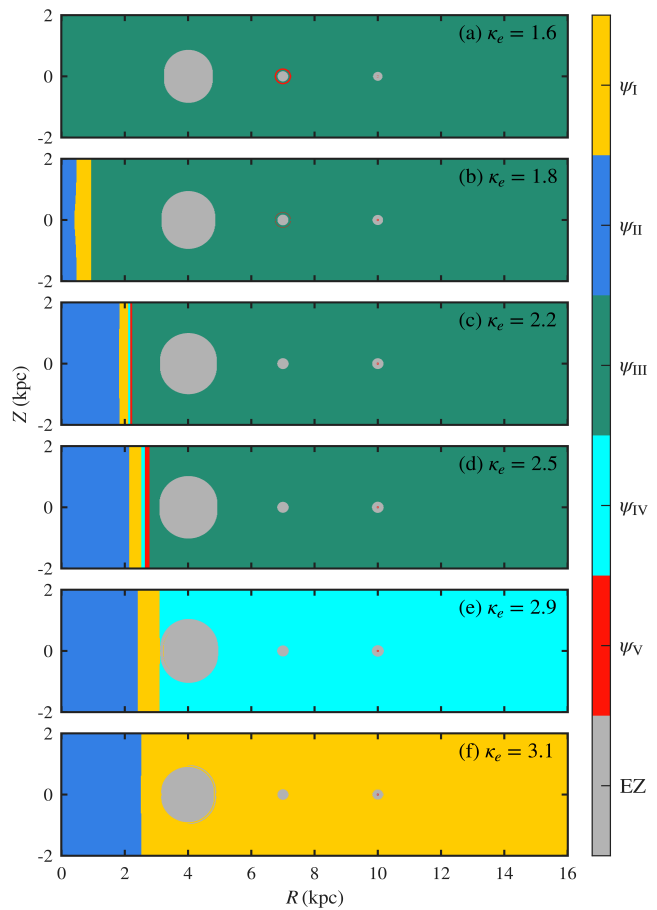}
\caption{Morphology-class maps of higher-order KA solitons in the ($R,Z$) plane for different $\kappa_e$. The centers of H\,\textsc{ii} region, SWB and SNR are located at $(R_1,Z_1)=(4,0)$, $(R_2,Z_2)=(7,0)$ and $(R_3,Z_3)=(10,0)$, respectively. Colors denote the localized solution class obtained at each grid point after applying the admissibility, regularity, and localization criteria. Gray regions indicate EZs where no admissible higher-order soliton exists.}
\label{fig:2d_morphology}
\end{figure}

Figure~\ref{fig:2d_morphology} presents the morphology maps across the $(R, Z)$ plane for six representative values of $\kappa_e$ used in Fig.~\ref{fig:1d_profiles}. The most striking feature of this figure is that the solution space is not described by a single continuous family that merely changes amplitude or width as one moves through the Galaxy, instead, the solution undergoes morphological transitions that are controlled jointly by the local ISM environment and by the degree of electron suprathermality.

At $\kappa_e = 1.6$ (Fig.~\ref{fig:2d_morphology}a), corresponding to the most strongly suprathermal electron population considered, the admissible region outside the exclusion zones is overwhelmingly dominated by $\psi_{\rm III}$. This result demonstrates that the plasma in the ISM, with a strong suprathermality of electrons, tends to support rarefactive, higher-order KA solitons with bifurcated topologies over the compressive single-peaked solitons predicted by first-order KdV theory \citep{SINGHInterstellar2026}.
The physical origin of this preference lies in the relative strengthening of higher-order contributions to the total potential in the strongly suprathermal regime. As $\kappa_e$ decreases, the coefficients of the evolution equation shift such that the leading-order nonlinear term becomes less dominant, while the second-order correction $\psi^{(2)}$ remains significant. Consequently, the total potential $\psi(\eta)$ reflects a nontrivial superposition of $\mathrm{sech}^2$ and $\mathrm{sech}^4$ components. When these contributions become comparable in magnitude, the solution departs from a single-peaked KdV profile and develops off-center extrema, giving rise to the observed double-hump morphology. 
The EZs, appearing as gray contours in Fig.~\ref{fig:2d_morphology}, are concentrated at the interiors of the embedded ISM structures (the H~\textsc{ii} region, the SWB, and the SNR cavity). The boundaries of these zones are physically meaningful: they do not indicate regions where no morphology was searched, but regions where no physically admissible higher-order localized solution exists under the current parameter constraints. 
Approaching these boundaries from the outside, the morphology undergoes rapid transitions, a consequence of the steep gradients in $\beta$ and density imposed by the ISM model at the edges of localized structures.

As $\kappa_e$ is increased to 1.8 (Fig.~\ref{fig:2d_morphology}b), the negative double-hump morphology $\psi_{\rm III}$ remains the primary feature across the outer Galactic disk, but a distinct structural evolution appears near the inner Galaxy. A localized region at $R \lesssim 0.5$ kpc develops a positive double-hump ($\psi_{\rm II}$) morphology, identified by the blue region. This indicates that softening the suprathermal tail and moving toward a more thermal electron distribution does not uniformly weaken the higher-order effects. Instead, the soliton experiences a polarity bifurcation where the central potential reverses sign while the double-peaked morphology is maintained. This confirms that the double-hump structure is a robust physical feature, dictated by the interaction between first and second-order perturbation terms, rather than a simple artifact of sign conventions. Furthermore, moving outward from this inner zone, the morphology transitions into a positive single-hump state ($\psi_{\rm I}$) between $0.5 \lesssim R \lesssim 1$\,kpc, represented by the yellow region. 
This spatial layering demonstrates how the high‑gradient environment of the inner Galaxy affects the higher‑order states before the wave reaches the more uniform regions of the diffuse medium.

As the electron population continues to thermalize toward $\kappa_e = 2.2$ and $2.5$ (Fig.~\ref{fig:2d_morphology}c and \ref{fig:2d_morphology}d), the inner Galactic region reveals a remarkably layered and radially dependent morphological sequence. 
At $\kappa_e = 2.2$, the positive double‑hump $\psi_{\rm II}$ (blue) occupies $R \lesssim 2$ kpc, then transitions to $\psi_{\rm I}$ (yellow), followed by the dressed soliton $\psi_{\rm IV}$ (cyan) and a narrow $\psi_{\rm V}$ (red) zone. Beyond that, the negative double‑hump $\psi_{\rm III}$ (green) dominates the rest of the ISM.
By $\kappa_e = 2.5$, this layered radial sequence persists but undergoes a finite spatial expansion. In this panel, a distinct widening is observed across all constituent regions, including the $\psi_{\rm I}$, $\psi_{\rm II}$, $\psi_{\rm IV}$, and $\psi_{\rm V}$. This global expansion suggests that as the suprathermal tail softens, the specific combination of plasma parameters necessary to sustain these diverse nonlinear states becomes available over a much larger radial extent of the inner disk. These panels provide the clearest evidence of the structured ISM functioning as a morphology regulator. While the suprathermality parameter $\kappa_e$ is held constant across the Galactic plane, the local coefficients governing nonlinearity and dispersion vary significantly with position through the structured model. This spatial variation forces the same wave mode into fundamentally different nonlinear states at different locations. The existence of the dressed state $\psi_{\rm IV}$ is particularly illustrative; it originates from the inhomogeneous forcing term $G(\psi^{(1)})$ in our evolution equation, which allows the second-order correction to fundamentally decorate the leading-order pulse with side lobes of opposite polarity. 
The simultaneous coexistence of the different morphological classes prove that the Galactic environment does not merely scale the wave amplitude, but actively determines the structural identity of the soliton by weighting the competition between leading and higher-order nonlinear terms.

A particularly important transition occurs at $\kappa_e = 2.9$ (Fig.~\ref{fig:2d_morphology}e). In this regime, the dressed soliton morphology $\psi_{\rm IV}$ (cyan) expands to become a dominant admissible state across a vast radial domain. This expansion represents one of the most significant findings of our analysis, identifying an intermediate suprathermal window where higher-order dressing is maximally expressed. While the inner Galactic regions remain characterized by the positive double-hump $\psi_{\rm II}$ (blue) and positive single-hump $\psi_{\rm I}$ (yellow) states. The widespread emergence of the $\psi_{\rm IV}$ class signifies that the dressed soliton is the typical nonlinear state of the medium under these conditions rather than an exotic or isolated solution. Physically, at $\kappa_e \approx 2.9$, the leading-order compressive tendency of the KdV core remains intact at the center of the profile; however, the higher-order dispersive and nonlinear balance is sufficiently strong to generate distinct rarefactive side structures. This interaction produces a genuinely composite waveform where the core and the dressing are of comparable significance. It is precisely this physical state for which the term \emph{dressed soliton} is most empirically justified, as the second-order corrections do not merely perturb the primary pulse but fundamentally redefine its structural identity.

By $\kappa_e = 3.1$ (Fig.~\ref{fig:2d_morphology}f), the global character of the morphology map reveals a pronounced simplification as the system approaches the thermal limit. The positive single-hump state $\psi_{\rm I}$, represented by the yellow region, now occupies the vast majority of the admissible Galactic plane. This indicates that for sufficiently large $\kappa_e$, higher-order corrections no longer produce significant topological bifurcations across the bulk of the ISM; instead, the nonlinear balance relaxes toward a KdV-like compressive state. Consequently, the high-$\kappa_e$ regime is characterized not merely by broader solitons, but by a distinct reduction in morphological variety. The positive double-hump morphology $\psi_{\rm II}$, shown in blue, remains restricted to the high-gradient Galactic region ($R \lesssim 2.5$ kpc), while the mixed-polarity and rarefactive double-hump branches disappear from the map. This spatial distribution results in a Galactic disk where the structural variety of solitons is strictly limited to the $\psi_{\rm I}$ and $\psi_{\rm II}$ states, signifying the restoration of leading-order dominance in the diffuse medium.

A notable feature of the morphology maps is the appearance of localized $\psi_{\rm V}$ structures in the immediate vicinity of SWB and inside SNR. At the SWB shell, a distinct red ring of $\psi_{\rm V}$ solitons encircles the bubble (most clearly seen in Figs.~\ref{fig:2d_morphology}a,b). This ring indicates that the split/degenerate double-hump state is preferentially excited at the shock-compressed boundary, where sharp gradients in density and magnetic field amplify the higher-order correction terms. As $\kappa_e$ increases, the ring gradually fades and disappears in the thermalised limit (Fig.~\ref{fig:2d_morphology}f). In contrast, the SNR exhibits a compact red $\psi_{\rm V}$ region near its centre ($R \approx 10$ kpc), embedded within the complex layered structure of the remnant. No such localized $\psi_{\rm V}$ features are found within the H\,{\sc ii} region, where the interior is either an exclusion zone or supports other morphological classes. These spatially confined higher-order states demonstrate that the embedded ISM structures are not passive background features but actively generate distinct soliton morphologies that would be absent in a homogeneous medium.

\subsection{Amplitude Diagnostics Along the Galactic Midplane}

\begin{figure}[!htbp]
\centering
\includegraphics[width=0.47\textwidth,height=0.67\textwidth]{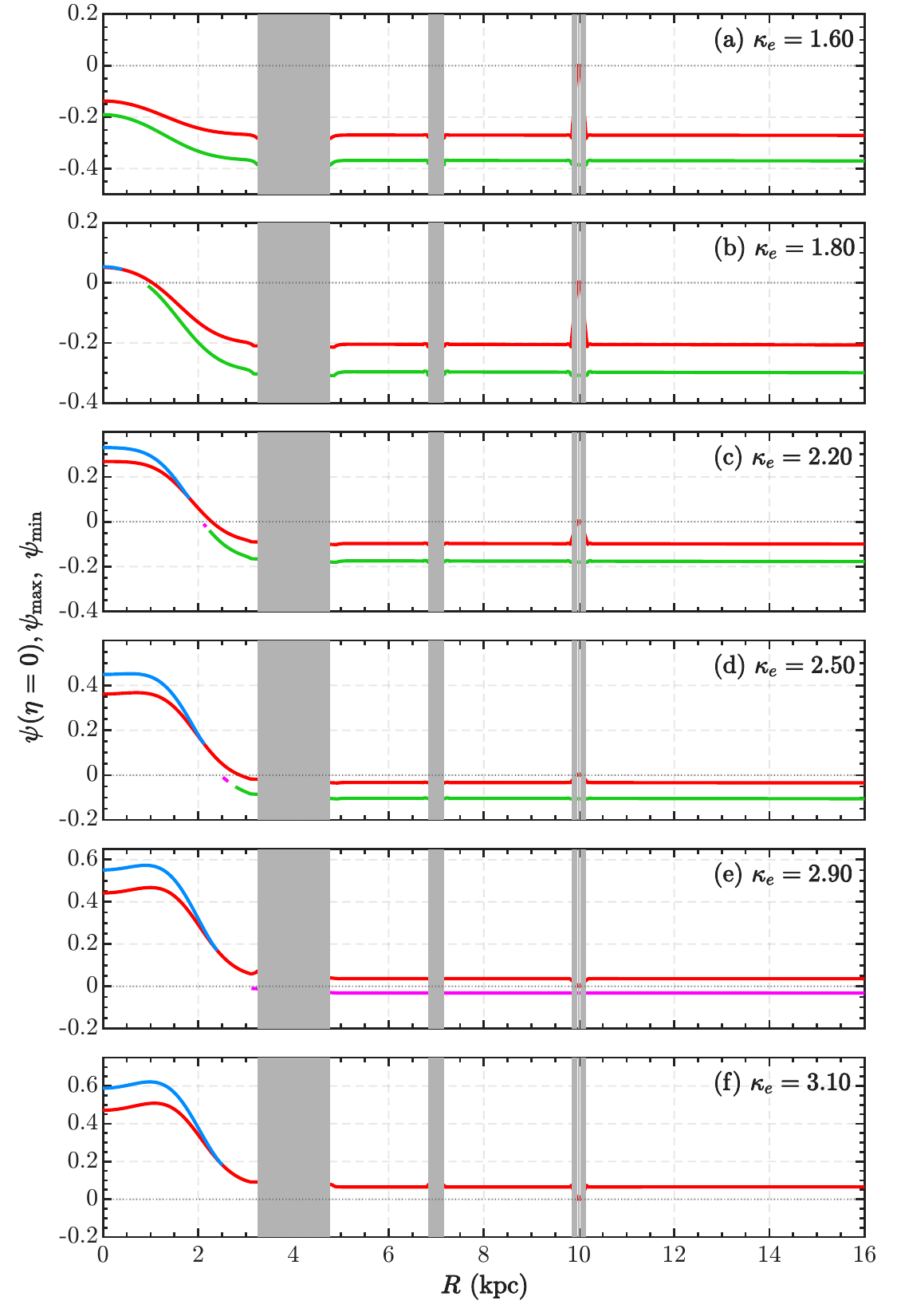}
\caption{Radial diagnostics at $Z=0$ showing the signed central potential $\psi(\eta=0)$ and the class-defining signed extrema for each $\kappa_e$. Blue curves denote positive double-hump peaks, green curves denote negative double-hump minima, and magenta curves denote negative-lobe minima of mixed-polarity dressed states. Gray shaded regions mark EZs where $m_e/m_i<\beta<1$ is not satisfied.}
\label{fig:1d_amplitudes}
\end{figure}

Figure~\ref{fig:1d_amplitudes} provides the quantitative analysis that complements the morphological map presented in Fig.~\ref{fig:2d_morphology}. By tracing the characteristic potential extrema as a function of the $R$ at the midplane ($Z = 0$), these curves act as a diagnostic bridge between the 2D spatial distributions and the representative waveform shapes. The red line tracks the central core potential $\psi(0)$, while auxiliary curves identify class-specific features: blue denotes the secondary peak $\psi_{\max}$ for positive double-hump states ($\psi_{\rm II}$), green tracks the secondary trough $\psi_{\min}$ for negative double-hump states ($\psi_{\rm III}$), and magenta identifies the negative-lobe minimum $\psi_{\min}$ characteristic of the dressed class ($\psi_{\rm IV}$). Gray shaded regions mark the EZs where the plasma $\beta$ exceeds the stability threshold.

For the strongly suprathermal case of $\kappa_e = 1.6$ (Fig.~\ref{fig:1d_amplitudes}a), both $\psi(0)$ and $\psi_{\min}$ remain negative across the entire disk, confirming the global dominance of the $\psi_{\rm III}$ morphology observed in Fig.~\ref{fig:2d_morphology}a. The comparable magnitudes of $\psi(0)$ and $\psi_{\min}$ indicate that higher-order corrections are sufficiently strong to maintain a bifurcated profile throughout the diffuse medium. As $\kappa_e$ increases to 1.8 (Fig.~\ref{fig:1d_amplitudes}b), the curves reveal the localized structural transitions noted in Fig.~\ref{fig:2d_morphology}b; specifically, the emergence of a blue curve representative of $\psi_{\rm II}$ (where both $\psi_{\max}$ and $\psi(0)$ positive) for $R \lesssim 0.5$ kpc and a lone positive red curve ($\psi_{\rm I}$) for $0.5 \lesssim R \lesssim 1$ kpc.

The midplane diagnostics for $\kappa_e = 2.2$ and 2.5 (Figs.~\ref{fig:1d_amplitudes}c and \ref{fig:1d_amplitudes}d) quantitatively resolve the radially layered sequence of morphological classes. In the inner Galactic region, the traces capture a clear transition from the positive double-hump class $\psi_{\rm II}$, where both the red $\psi(0)$ and blue $\psi_{\max}$ curves are positive, into the positive single-hump state $\psi_{\rm I}$, characterized by the lone presence of the positive red curve. This is followed by the emergence of the dressed soliton class $\psi_{\rm IV}$, identified by the appearance of magenta $\psi_{\min}$ patches alongside the positive $\psi(0)$ red line. Further toward the outer disk, the diagnostics reveal a re-emergence of the negative double-hump structures $\psi_{\rm III}$, where both $\psi(0)$ and the green $\psi_{\min}$ curves remain negative. Consistent with the spatial expansion observed in Figure~\ref{fig:2d_morphology}d, all diagnostic traces broaden significantly at $\kappa_e = 2.5$; this expansion indicates that thermalization increases the radial existence window for these higher-order states. In this regime, the prominence of the magenta signatures highlights the $\psi_{\rm IV}$ domain, where higher-order dressing generates the rarefactive side lobes that decorate the compressive core.

An important transition occurs at $\kappa_e = 2.9$ (Fig.~\ref{fig:1d_amplitudes}e), where the dressed soliton $\psi_{\rm IV}$ become the dominant morphology across much of the midplane. The persistence of the magenta $\psi_{\min}$ curve alongside a positive $\psi(0)$ over nearly the entire Galactic disk (for $R\gtrsim 3$ kpc) validates the global dominance of the green region in Fig.~\ref{fig:2d_morphology}e. Finally, at $\kappa_e = 3.1$ (Fig.~\ref{fig:1d_amplitudes}f), the diagnostics reveal morphological simplification; the auxiliary lines vanish over the bulk of the disk (for $R\gtrsim 2.5$ kpc), leaving only the positive core potential $\psi(0)$. This lone red trace confirms the relaxation into the positive single-hump $\psi_{\rm I}$ morphology, with the $\psi_{\rm II}$ state (blue curve) still sustaining in the high-gradient region $R \lesssim 2.5$ kpc, in exact agreement with the simplified map of Figure~\ref{fig:2d_morphology}f.

The radial diagnostics (Fig.~\ref{fig:1d_amplitudes}) show how soliton morphology responds to EZs (gray). At the H\,{\sc ii} region, $\psi(0)$ and auxiliary curves decrease only slightly at the EZ edge due to the smooth Gaussian boundary. Inside the SWB cavity, $\psi(0)$ is suppressed, but at the shell boundaries auxiliary traces show large changes (most pronounced in panels~a,b), corresponding to the red $\psi_{\mathrm{V}}$ ring in Fig.~\ref{fig:2d_morphology}, which show a brief amplification of higher-order components before the soliton vanishes. For the SNR, the central PWN core is a gray EZ where no soliton exists. Immediately outside this core, a non-EZ region appears (red curve in Fig.~\ref{fig:1d_amplitudes}), but the amplitude $\psi(0)$ is negligible (nearly zero) there,  these are not physically meaningful solitons. Further out, another EZ (gray) occurs, followed by the SNR shell (non-EZ), where auxiliary curves show only minor variations and vanish in the thermalised limit (panel~f). Meaningful higher-order dressing occurs only where $m_e/m_i<\beta<1$ and the amplitude is finite; inside EZs or where $\psi(0)\approx 0$, no admissible soliton exists.

\section{Summary and Conclusion}
\label{Summary}

We have extended the spatially dependent framework for KA solitons in the ISM to the higher‑order nonlinear regime. Using a fluid model with superthermal electrons and a structured Galactic environment (WIM, H\,\textsc{ii} regions, SWBs, SNRs), we derived an inhomogeneous KdV‑type equation that includes cubic nonlinearity, nonlinear‑dispersive cross terms, and fifth‑order dispersion. The closed‑form dressed soliton consists of a leading‑order $\operatorname{sech}^2$ core decorated by $\operatorname{sech}^2$ and $\operatorname{sech}^4$ corrections whose relative weights are set by the local ISM parameters.

We classified soliton morphologies ($\psi_{\rm I}$–$\psi_{\rm V}$) based on the central potential $\psi(0)$ and the dominant extrema, and mapped them across the Galactic plane. The distribution evolves non‑monotonically with the electron suprathermality index $\kappa_e$:
\begin{itemize}
    \item At $\kappa_e=1.6$ (strongly suprathermal), negative double‑hump solitons ($\psi_{\rm III}$) dominate the admissible regions.
    \item At $\kappa_e=1.8$, a polarity bifurcation creates a positive double‑hump ($\psi_{\rm II}$) region for $R\lesssim0.5$ kpc, followed by $\psi_{\rm I}$ for $0.5\lesssim R\lesssim1$ kpc. 
    Overall, different soliton classes appear in following sequence: $\psi_{\rm II}$ (blue) $\to$ $\psi_{\rm I}$ (yellow) $\to$ $\psi_{\rm III}$ (green).
    \item At $\kappa_e=2.2$–$2.5$, a radially layered sequence appears: $\psi_{\rm II}$ (blue) $\to$ $\psi_{\rm I}$ (yellow) $\to$ dressed $\psi_{\rm IV}$ (cyan) $\to$ $\psi_{\rm V}$ (red) $\to$ $\psi_{\rm III}$ (green).
    \item At $\kappa_e=2.9$, the dressed soliton $\psi_{\rm IV}$ expands to become the dominant state across a vast radial domain, marking an intermediate window where higher‑order dressing is maximally expressed. At this $\kappa_e$, different soliton classes appear in following sequence: $\psi_{\rm II}$ (blue) $\to$ $\psi_{\rm I}$ (yellow) $\to$ $\psi_{\rm IV}$ (cyan).
    \item At $\kappa_e=3.1$ (near‑Maxwellian), the system relaxes to positive single‑hump KdV‑like states ($\psi_{\rm I}$) over most of the disk, with only residual $\psi_{\rm II}$ in the inner Galaxy.
\end{itemize}
Localized $\psi_{\rm V}$ structures appear as a red ring around the SWB shell (most clearly in Figs.~\ref{fig:2d_morphology}a,b) and a compact red dot inside the SNR, demonstrating that embedded ISM structures actively generate distinct higher‑order soliton morphologies absent in a homogeneous medium. The radial amplitude diagnostics (Fig.~\ref{fig:1d_amplitudes}) show that at the H\,\textsc{ii}  region $\psi(0)$ decreases only slightly; at the SWB shell auxiliary traces exhibit large changes; and inside the SNR the central EZ (PWN core) gives way to a non‑EZ region with negligible amplitude, then another EZ, and finally the shell with minor variations.

Our main conclusions are: (i) First‑order KdV theory is quantitatively insufficient in large parts of the ISM (inner Galaxy, SWB/SNR shells); dressed solitons are the natural nonlinear states. (ii) The ISM morphology does not merely scale amplitudes but actively controls the soliton class by modulating the competition between leading‑order and higher‑order terms. (iii) Localised $\psi_{\rm V}$ features provide a direct link between macroscopic ISM structures (SWB shells, SNR interiors) and coherent kinetic‑scale fluctuations, offering candidates for extreme scattering events and pulsar scintillation. (iv) The non‑monotonic dependence on $\kappa_e$ opens a path to constrain the electron suprathermality from scintillation observations. This framework provides a model for galaxy‑scale simulations and can be extended to other plasma phenomena in structured astrophysical environments.

\begin{acknowledgements}
M.S. gratefully acknowledges support from the Basic Scientific Research Fund for Central Universities, China (Grant No. 2682025CX094). S.L. acknowledges funding from the National Natural Science Foundation of China (Grant No. 12375103). N.S.S. acknowledges support from the Council of Scientific and Industrial Research (India) under the Emeritus Scientist scheme (Ref. No. 21/1195/25/EMR‑II).
\end{acknowledgements}

\bibliographystyle{aa}
\bibliography{Ref}


\appendix

\begin{table*}[!h]
\caption{Parameters of embedded ISM structures. Amplitudes $A<0$ indicate depletion. Lengths in kpc, $n_e$ in cm$^{-3}$, $B$ in $\mu$G, $T$ in K. Central coordinates: H\,II: $(R_1,Z_1)$, SWB: $(R_2,Z_2)$, SNR: $(R_3,Z_3)$. Table adopted from \citet{SINGHInterstellar2026}.}
\label{tab:profiles_parameters}
\centering
\setlength{\tabcolsep}{4pt}
\begin{tabular}{llp{0.65\textwidth}}
\hline
\hline
Structure & Quantity & Parameters \\
\hline
H\,II Region
& $n_{e,\mathrm{HII}}$
& $A_{n,\mathrm{HII}}=\text{2000 $\times$ $n_{e,\mathrm{bkg}}(R_1,Z_1)$}$, $\sigma_{n,\mathrm{HII}}=0.25$. \\
&  $B_{0,\mathrm{HII}}$
& $A_{B,\mathrm{HII}}=\texttt{$-0.7\times B_{0,\mathrm{bkg}}(R_1,Z_1)$}$, $\sigma_{B,\mathrm{HII}}=0.25$. \\
&  $T_{\mathrm{HII}}$
& $A_{T,\mathrm{HII}}= 0.25\,\times\,T_{e,\text{bkg}}$, $\sigma_{T,\mathrm{HII}}=0.25$. \\
\hline
SWB Shell
& $n_{e,\mathrm{sh,SWB}}$
& $A_{n,\mathrm{sh,SWB}}=\texttt{$4\times n_{e,\mathrm{bkg}}(R_2,Z_2)$}$, $R_{\mathrm{sh,SWB}}=\texttt{0.25}$\newline
  $\sigma_{n,\mathrm{in,SWB}}=\texttt{$R_{\mathrm{sh,SWB}}/10$}$, $\sigma_{n,\mathrm{out,SWB}}=\texttt{$\sigma_{n,\mathrm{in,SWB}}/4$}$. \\
&  $B_{0,\mathrm{sh,SWB}}$
& $A_{B,\mathrm{sh,SWB}}=\texttt{$10\times B_{0,\mathrm{bkg}}(R_3,Z_3)$}$, $\sigma_{B,\mathrm{in,SWB}}=\texttt{$R_{\mathrm{sh,SWB}}/10$}$\newline
  $\sigma_{B,\mathrm{out,SWB}}=\texttt{$\sigma_{B,\mathrm{in,SWB}}/4$}$. \\
&  $T_{e,\mathrm{sh,SWB}}$
& $A_{T,\mathrm{sh,SWB}}=\texttt{$10^4\,$K$-T_{e,\mathrm{bkg}}$}$, $\sigma_{T,\mathrm{in,SWB}}=\texttt{$R_{\mathrm{sh,SWB}}/2$}$\newline
  $\sigma_{T,\mathrm{out,SWB}}=\texttt{$\sigma_{n,\mathrm{in,SWB}}/6$}$. \\
\cline{2-3}
SWB Cavity
&  $n_{e,\mathrm{cav,SWB}}$
& $A_{n,\mathrm{cav,SWB}}=\texttt{$-0.9\times n_{e,\mathrm{bkg}}(R_2,Z_2)$}$, $\sigma_{n,\mathrm{cav,SWB}}=\texttt{$R_{\mathrm{sh,SWB}}/2$}$. \\
&  $B_{0,\mathrm{cav,SWB}}$
& $A_{B,\mathrm{cav,SWB}}=\texttt{$-0.5\times B_{0,\mathrm{bkg}}(R_2,Z_2)$}$, $\sigma_{B,\mathrm{cav,SWB}}=\texttt{$R_{\mathrm{sh,SWB}}/2$}$. \\
\cline{2-3}
SWB Interior
&  $T_{e,\mathrm{plat,SWB}}$
& $A_{T,\mathrm{plat,SWB}}=\texttt{$10^6\,$K$-T_{e,\mathrm{bkg}}$}$, $R_{\mathrm{flat,SWB}}=\texttt{$0.6\times R_{\mathrm{sh,SWB}}$}$. \\
\hline
SNR Shell
&  $n_{e,\mathrm{sh,SNR}}$
& $A_{n,\mathrm{sh,SNR}}=\texttt{$4\times n_{e,\mathrm{bkg}}(R_3,Z_3)$}$, $R_{\mathrm{sh,SNR}}=0.25$\newline
  $\sigma_{n,\mathrm{in,SNR}}=\texttt{$R_{\mathrm{sh,SNR}}/10$}$, $\sigma_{n,\mathrm{out,SNR}}=\texttt{$\sigma_{n,\mathrm{in,SNR}}/10$}$. \\
& $B_{0,\mathrm{sh,SNR}}$
& $A_{B,\mathrm{sh,SNR}}=\texttt{$4\times B_{0,\mathrm{bkg}}(R_3,Z_3)$}$, $\sigma_{B,\mathrm{in,SNR}}=\texttt{$R_{\mathrm{sh,SNR}}/10$}$\newline
  $\sigma_{B,\mathrm{out,SNR}}=\texttt{$\sigma_{B,\mathrm{in,SNR}}/10$}$. \\
& $T_{e,\mathrm{sh,SNR}}$
& $A_{T,\mathrm{sh,SNR}}=\texttt{$(10^4\,$K$-T_{e,\mathrm{bkg}})$}$, $\sigma_{T,\mathrm{in,SNR}}=\texttt{$R_{\mathrm{sh,SNR}}/1.5$}$\newline
  $\sigma_{T,\mathrm{out,SNR}}=\texttt{$\sigma_{T,\mathrm{in,SNR}}/10$}$. \\
\cline{2-3}
SNR Cavity
&  $n_{e,\mathrm{cav,SNR}}$
& $A_{n,\mathrm{cav,SNR}}=\texttt{$-\,n_{e,\mathrm{bkg}}(R_3,Z_3)$}$, $\sigma_{n,\mathrm{cav,SNR}}=\texttt{$R_{\mathrm{sh,SNR}}/3$}$. \\
\cline{2-3}
SNR PWN
&  $n_{e,\mathrm{PWN,SNR}}$
& $A_{n,\mathrm{PWN,SNR}}=\texttt{$n_{e,\mathrm{bkg}}(R_3,Z_3)/1.25$}$, $\sigma_{n,\mathrm{PWN,SNR}}=\texttt{$R_{\mathrm{sh,SNR}}/10$}$. \\
&  $B_{0,\mathrm{PWN,SNR}}$
& $A_{B,\mathrm{PWN,SNR}}=\texttt{120}\,\mu\mathrm{G}$, $\sigma_{B,\mathrm{PWN,SNR}}=\texttt{$R_{\mathrm{sh,SNR}}/25$}$. \\
&  $p_{\mathrm{SNR}}$
& $A_{p,\mathrm{SNR}}=\texttt{$(p_{\mathrm{peak,SNR}}-p_{\mathrm{bkg}})$}$, with $\texttt{$p_{\mathrm{peak,SNR}}=10^4$}$\newline
  $\sigma_{p,\mathrm{SNR}}=\texttt{$R_{\mathrm{sh,SNR}}/25$}$. \\
&  $T_{e,\mathrm{core,SNR}}$
& $A_{T,\mathrm{core,SNR}}=\texttt{$(10^7+T_{e,\mathrm{bkg}})$}$, $\sigma_{T,\mathrm{core,SNR}}=\texttt{$R_{\mathrm{sh,SNR}}/5$}$. \\
\hline
\hline
\end{tabular}
\end{table*}

\section{First-Order Perturbation Equations}
\label{app:FO}

Substituting the RPM stretching (\ref{eq8}), the power-series expansions (\ref{eq10})--(\ref{eq11}), and the ion density expansion (\ref{eq12}) into the fluid equations~(\ref{eq1})--(\ref{eq4}) and retaining terms at $\mathcal{O}(\epsilon^{3/2})$ yields the following system of four coupled equations relating the first-order perturbed fields $v_{ix}^{(1)}$, $v_{iz}^{(1)}$, $\psi^{(1)}$, and $\phi^{(1)}$:
\begin{equation}
-a_{1}\lambda\frac{\partial\psi^{(1)}}{\partial\xi}
  +l_{x}\frac{\partial v_{ix}^{(1)}}{\partial\xi}
  +l_{z}\frac{\partial v_{iz}^{(1)}}{\partial\xi}=0,
\label{appA:eq1}
\end{equation}
\begin{equation}
v_{ix}^{(1)}=\Lambda\lambda l_{x}
  \frac{\partial^{2}\phi^{(1)}}{\partial\xi^{2}},
\label{appA:eq2}
\end{equation}
\begin{equation}
\lambda\frac{\partial v_{iz}^{(1)}}{\partial\xi}
  =\Lambda l_{z}\frac{\partial\psi^{(1)}}{\partial\xi},
\label{appA:eq3}
\end{equation}
\begin{equation}
\Lambda l_{x}^{2}l_{z}^{2}
  \frac{\partial^{2}\phi^{(1)}}{\partial\xi^{2}}
  =a_{1}\lambda^{2}\psi^{(1)}-l_{z}\lambda v_{iz}^{(1)}.
\label{appA:eq4}
\end{equation}
Algebraically eliminating $v_{ix}^{(1)}$, $v_{iz}^{(1)}$, and $\phi^{(1)}$ from Eqs.~(\ref{appA:eq1})--(\ref{appA:eq4}) yields the biquadratic dispersion relation Eq.~(\ref{eq14}) of the main text, whose physical roots are given in Eqs.~(\ref{eq15})--(\ref{eq16}).  On the KA branch ($\lambda^2=l_z^2$), the first-order solutions read
\begin{equation}
v_{iz}^{(1)}=\Lambda\psi^{(1)},\qquad
\phi^{(1)}=\frac{a_1\lambda}{l_x^2 l_z^2 \Lambda}\psi^{(1)}.
\label{appA:sol}
\end{equation}
These expressions are used as known driving terms when constructing the second-order system in Appendix~\ref{app:SO}.

\section{Second-Order Perturbation Equations}
\label{app:SO}

Retaining terms at $\mathcal{O}(\epsilon^{5/2})$ in Eqs.~(\ref{eq1})--(\ref{eq4}) yields four equations coupling the second-order quantities $v_{ix}^{(2)}$, $v_{iz}^{(2)}$, $\psi^{(2)}$, and $\phi^{(2)}$ to nonlinear products of the first-order fields:
\begin{eqnarray}
-a_{1}\lambda\frac{\partial\psi^{(2)}}{\partial\xi}
  -2a_{2}\lambda\psi^{(1)}\frac{\partial\psi^{(1)}}{\partial\xi}
  +a_{1}\frac{\partial\psi^{(1)}}{\partial\tau}
  +l_{x}\frac{\partial v_{ix}^{(2)}}{\partial\xi}
\nonumber\\
+\,a_{1}l_{x}\frac{\partial(\psi^{(1)}v_{ix}^{(1)})}{\partial\xi}
  +l_{z}\frac{\partial v_{iz}^{(2)}}{\partial\xi}
  +a_{1}l_{z}\frac{\partial(\psi^{(1)}v_{iz}^{(1)})}{\partial\xi}=0,
\label{appB:eq1}
\end{eqnarray}
\begin{equation}
v_{ix}^{(2)}=\Lambda\!\left(
  \lambda l_{x}\frac{\partial^{2}\phi^{(2)}}{\partial\xi^{2}}
  -l_{x}\frac{\partial^{2}\phi^{(1)}}{\partial\tau\partial\xi}
  \right),
\label{appB:eq2}
\end{equation}
\begin{equation}
-\lambda\frac{\partial v_{iz}^{(2)}}{\partial\xi}
  +\frac{\partial v_{iz}^{(1)}}{\partial\tau}
  +l_{x}v_{ix}^{(1)}\frac{\partial v_{iz}^{(1)}}{\partial\xi}
  +l_{z}v_{iz}^{(1)}\frac{\partial v_{iz}^{(1)}}{\partial\xi}
  =-\Lambda l_{z}\frac{\partial^{2}\psi^{(2)}}{\partial\xi^{2}},
\label{appB:eq3}
\end{equation}
\begin{eqnarray}
\Lambda l_{x}^{2}l_{z}^{2}
  \frac{\partial^{4}(\phi^{(2)}-\psi^{(1)})}{\partial\xi^{4}}
  =a_{1}\lambda^{2}\frac{\partial^{2}\psi^{(2)}}{\partial\xi^{2}}
  +a_{2}\lambda^{2}\frac{\partial^{2}\psi^{(1)}}{\partial\xi^{2}}
\nonumber\\
-\,2a_{1}\lambda\frac{\partial^{2}\psi^{(1)}}{\partial\xi\partial\tau}
  -l_{z}\lambda\frac{\partial^{2}v_{iz}^{(2)}}{\partial\xi^{2}}
  +l_{z}\frac{\partial^{2}v_{iz}^{(1)}}{\partial\xi\partial\tau}
\nonumber\\
-\,a_{1}l_{z}\lambda
  \frac{\partial^{2}(v_{iz}^{(1)}\psi^{(1)})}{\partial\xi^{2}}.
\label{appB:eq4}
\end{eqnarray}
Eliminating $v_{ix}^{(2)}$, $v_{iz}^{(2)}$, and $\phi^{(2)}$ from Eqs.~(\ref{appB:eq1})--(\ref{appB:eq4}) using the first-order solutions (\ref{appA:sol}) yields the KdV equation~(\ref{eq17}) with coefficients $A$ and $B$ given in Eqs.~(\ref{eq18})--(\ref{eq19}). The second-order quantities $v_{iz}^{(2)}$ and $\phi^{(2)}$ obtained from Eqs.~(\ref{appB:eq3})--(\ref{appB:eq4}) enter as additional driving terms in the third-order system of Appendix~\ref{app:TO}.

\section{Third-Order Perturbation Equations}
\label{app:TO}

Retaining terms at $\mathcal{O}(\epsilon^{7/2})$ in Eqs.~(\ref{eq1})--(\ref{eq4}) yields four equations coupling the third-order quantities $v_{ix}^{(3)}$, $v_{iz}^{(3)}$, $\psi^{(3)}$, and $\phi^{(3)}$ to products of first- and second-order fields:
\begin{eqnarray}
-a_{1}\lambda\frac{\partial\psi^{(3)}}{\partial\xi}
  -2a_{2}\lambda\!\left(\psi^{(1)}\frac{\partial\psi^{(2)}}{\partial\xi}
                +\psi^{(2)}\frac{\partial\psi^{(1)}}{\partial\xi}\right)
  -3a_{3}\lambda{\psi^{(1)}}^{2}\frac{\partial\psi^{(1)}}{\partial\xi}
\nonumber\\
+\,a_{1}\frac{\partial\psi^{(2)}}{\partial\tau}
  +l_{x}\frac{\partial v_{ix}^{(3)}}{\partial\xi}
  +a_{1}l_{x}\frac{\partial(\psi^{(1)}v_{ix}^{(2)}
                            +\psi^{(2)}v_{ix}^{(1)})}{\partial\xi}
\nonumber\\
+\,l_{z}\frac{\partial v_{iz}^{(3)}}{\partial\xi}
  +a_{1}l_{z}\frac{\partial(\psi^{(1)}v_{iz}^{(2)}
                            +\psi^{(2)}v_{iz}^{(1)})}{\partial\xi}=0,
\label{appC:eq1}
\end{eqnarray}
\begin{equation}
v_{ix}^{(3)}=\Lambda\!\left(
  \lambda l_{x}\frac{\partial^{2}\phi^{(3)}}{\partial\xi^{2}}
  -l_{x}\frac{\partial^{2}\phi^{(2)}}{\partial\tau\partial\xi}
  \right),
\label{appC:eq2}
\end{equation}
\begin{eqnarray}
\lambda\frac{\partial v_{iz}^{(3)}}{\partial\xi}
  +\frac{\partial v_{iz}^{(2)}}{\partial\tau}
  +l_{x}\!\left(v_{ix}^{(1)}\frac{\partial v_{iz}^{(2)}}{\partial\xi}
              +v_{ix}^{(2)}\frac{\partial v_{iz}^{(1)}}{\partial\xi}\right)
\nonumber\\
+\,l_{z}\!\left(v_{iz}^{(1)}\frac{\partial v_{iz}^{(2)}}{\partial\xi}
              +v_{iz}^{(2)}\frac{\partial v_{iz}^{(1)}}{\partial\xi}\right)
  =-\Lambda l_{z}\frac{\partial^{2}\psi^{(3)}}{\partial\xi^{2}},
\label{appC:eq3}
\end{eqnarray}
\begin{eqnarray}
\Lambda l_{x}^{2}l_{z}^{2}
  \frac{\partial^{4}(\phi^{(3)}-\psi^{(2)})}{\partial\xi^{4}}
  =a_{1}\lambda^{2}\frac{\partial^{2}\psi^{(3)}}{\partial\xi^{2}}
\nonumber\\
+\,a_{2}\lambda^{2}
  \frac{\partial^{2}(\psi^{(1)}\psi^{(2)})}{\partial\xi^{2}}
  +a_{3}\lambda^{2}\frac{\partial^{2}{\psi^{(1)}}^{3}}{\partial\xi^{2}}
\nonumber\\
-\,2a_{1}\lambda\frac{\partial^{2}\psi^{(2)}}{\partial\xi\partial\tau}
  -l_{z}\lambda\frac{\partial^{2}v_{iz}^{(3)}}{\partial\xi^{2}}
  +l_{z}\frac{\partial^{2}v_{iz}^{(2)}}{\partial\xi\partial\tau}
\nonumber\\
-\,a_{1}l_{z}\lambda
  \frac{\partial^{2}(v_{iz}^{(1)}\psi^{(2)}
                    +v_{iz}^{(2)}\psi^{(1)})}{\partial\xi^{2}}.
\label{appC:eq4}
\end{eqnarray}
Here $a_3$ is the third-order coefficient in the Taylor expansion of the ion density (\ref{eq12}).  Eliminating $v_{ix}^{(3)}$, $v_{iz}^{(3)}$, $\psi^{(3)}$, and $\phi^{(3)}$ from Eqs.~(\ref{appC:eq1})--(\ref{appC:eq4}), substituting the first-order solutions from Appendix~\ref{app:FO} and the second-order solutions from Appendix~\ref{app:SO}, and applying the Fredholm-type solvability condition \citep{Kodama1978} yields the inhomogeneous equation~(\ref{eq23}) with forcing function $G(\psi^{(1)})$ given in Eq.~(\ref{G_psi}).

\section{Higher-Order Coefficients}
\label{app:coeffs}

The forcing function $G(\psi^{(1)})$ in Eq.~(\ref{G_psi}) depends on four coefficients $A_1$--$A_4$, which are determined entirely by the leading-order KdV coefficients ${\cal A}$, ${\cal B}$, and the equilibrium density expansion coefficients $a_1$, $a_2$.  They take the form
\begin{equation}
A_{1}=\frac{P}{{\cal Z}}, \qquad
A_{2}=\frac{{\cal Q}}{{\cal Z}}, \qquad
A_{3}=\frac{{\cal R}}{{\cal Z}}, \qquad
A_{4}=\frac{S}{{\cal Z}},
\label{appD:Ai}
\end{equation}
where
\begin{eqnarray}
P &=& -l_{x}^{2}\lambda^{2}a_{3}
     +\frac{l_{x}\lambda a_{1}}{2\Lambda}
     +\frac{5\lambda^{2}a_{1}a_{2}}{3\Lambda}
     -\frac{a_{1}^{2}\lambda^{2}}{3}
     -\frac{a_{1}\Lambda\lambda^{2}}{3}
\nonumber\\
  &&+\frac{a_{1}\Lambda\lambda^{3}}{3}
     +\frac{a_{1}\lambda^{2}}{\Lambda}
     +a_{1}^{3}\lambda^{2}
     -\frac{a_{2}\lambda^{2}}{3}
     +\frac{a_{3}\lambda^{2}}{\Lambda},
\label{appD:P}
\end{eqnarray}
\begin{equation}
{\cal Q} = 7{\cal AB}
    -\frac{4a_{1}^{2}{\cal B}\lambda}{\Lambda}
    +4a_{1}{\cal B}\lambda
    +\frac{4a_{2}{\cal B}\lambda}{\Lambda}
    -\frac{\Lambda a_{1}{\cal B}}{\lambda},
\label{appD:Q}
\end{equation}
\begin{equation}
{\cal R} = -2{\cal AB}
    +\frac{3a_{1}^{2}{\cal B}\lambda}{\Lambda}
    +3a_{1}{\cal B}\lambda
    -\frac{4a_{2}{\cal B}\lambda}{\Lambda}
    -2\Lambda {\cal B}\lambda,
\label{appD:R}
\end{equation}
\begin{equation}
S = 3{\cal B}^{2}+\frac{a_{1}{\cal B}^{2}}{\Lambda},
\label{appD:S}
\end{equation}
\begin{equation}
{\cal Z} = \frac{2\lambda(a_{1}-\Lambda)}{\Lambda}.
\label{appD:Z}
\end{equation}
The quantities $\lambda$, ${\cal A}$, and ${\cal B}$ are given by Eqs.~(\ref{eq15}), (\ref{eq18}), and (\ref{eq19}), respectively.  Since all of these depend on the local ISM parameters at $(R,Z)$, the coefficients $A_1$--$A_4$ inherit the full spatial dependence of the structured ISM model.  In particular, the sign and magnitude of $A_4$ determine the correction to the soliton speed $U$ in Eq.~(\ref{eq29}) and width $D$ in Eq.~(\ref{HO_width}), while $A_1$--$A_3$ regulate the relative weight of the nonlinear forcing contributions in $G(\psi^{(1)})$ that produce the off-center morphological features of the dressed profile.

\end{document}